\input harvmac.tex      
\input epsf.tex         


%
%
%

\def\tilde{\widetilde}
\def\bar{\overline}

\def\*{\star}
\def\[{\left[}
\def\]{\right]}
\def\({\left(}		
\def\){\right)}

%
%
\def\zb{{\bar{z} }}
\def\frac#1#2{{#1 \over #2}}
\def\inv#1{{1 \over #1}}

\def\d{\partial}

\def\2pi{\hbox{$2\pi i$}}

\def\dsl{\raise.15ex\hbox{/}\kern-.57em\partial}
\def\Dsl{\,\raise.15ex\hbox{/}\mkern-.13.5mu D}
%
%

%
%
	\def\CB{{\cal B}}	\def\CC{{\cal C}}
\def\CD{{\cal D}}		
		
		\def\CL{{\cal L}}
\def\CM{{\cal M}}		\def\CO{{\cal O}}
		\def\CR{{\cal R}}

\def\2pi{\hbox{$2\pi i$}}

\def\dsl{\raise.15ex\hbox{/}\kern-.57em\partial}
\def\Dsl{\,\raise.15ex\hbox{/}\mkern-.13.5mu D}
%
%
%
\font\numbers=cmss12
\font\upright=cmu10 scaled\magstep1
\def\stroke{\vrule height8pt width0.4pt depth-0.1pt}
\def\topfleck{\vrule height8pt width0.5pt depth-5.9pt}
\def\botfleck{\vrule height2pt width0.5pt depth0.1pt}
\def\Zmath{\vcenter{\hbox{\numbers\rlap{\rlap{Z}\kern
0.8pt\topfleck}\kern
2.2pt
                   \rlap Z\kern 6pt\botfleck\kern 1pt}}}
\def\Qmath{\vcenter{\hbox{\upright\rlap{\rlap{Q}\kern
                   3.8pt\stroke}\phantom{Q}}}}
\def\Nmath{\vcenter{\hbox{\upright\rlap{I}\kern 1.7pt N}}}
\def\Cmath{\vcenter{\hbox{\upright\rlap{\rlap{C}\kern
                   3.8pt\stroke}\phantom{C}}}}
\def\Rmath{\vcenter{\hbox{\upright\rlap{I}\kern 1.7pt R}}}
\def\Z{\ifmmode\Zmath\else$\Zmath$\fi}
\def\Q{\ifmmode\Qmath\else$\Qmath$\fi}
\def\N{\ifmmode\Nmath\else$\Nmath$\fi}
\def\C{\ifmmode\Cmath\else$\Cmath$\fi}
\def\R{\ifmmode\Rmath\else$\Rmath$\fi}

\Title{ITP-97-080, hep-th/9708135}
{\vbox{\centerline{Minimal Models with Integrable Local Defects }
\centerline{ ~} }}

\bigskip
\bigskip

\centerline{Andr\'e LeClair\foot{On leave from Cornell University,
Newman Laboratory, Ithaca, NY 14853.}${}^{,a}$
and Andreas  W.W. Ludwig${}^{a,b}$}
\bigskip
\medskip
\centerline{
${}^{a}$Institute for Theoretical Physics}
\centerline{University of California}
\centerline{Santa Barbara, CA 93106-4030}
\medskip
\medskip
\medskip\centerline{
${}^{b}$Department of Physics
}
\centerline{University of California}
\centerline{Santa Barbara, CA 93106-4030}

\vskip .3in

We describe a general way of constructing integrable defect theories
as perturbations of conformal field theory by local defect operators.
The method relies on folding the system onto a boundary field theory
 of {\it twice} the central charge.
The  classification of integrable  defect theories 
obtained in this way parallels that 
 of integrable bulk theories which are a perturbation
of the tensor product of {\it two} conformal field theories.
These include local defect perturbations of all
$c<1$ minimal models, as well as of the coset theories
based on $SO(2n)$, obtained in this way.
 We discuss  in detail the former case of all the Virasoro
minimal models.
In the Ising case our construction corresponds  to
 having a spin
field as a defect operator;  in the folded formulation this is
mapped onto an orbifolding of the boundary sine-Gordon theory at
$\beta^2/8\pi = 1/8$, or a version of the anisotropic Kondo model.

\Date{8/97}
%
%
%
%
%
%

%
%
%
%
%
%
%
%
%
%
%

\def\L#1{#1^{(L)}}
\def\R#1{#1^{(R)}}
\def\p#1{#1^{(+)}}
\def\m#1{#1^{(-)}}

\newsec{Introduction}

Two dimensional quantum field theories with impurities or defects
have received  a great deal  of attention in the past few years.  
In the $(1+1)$ quantum context, such theories 
have many important applications,  including
 `weak links' in infinite
$s=1/2$ Heisenberg Quantum Spin chains 
\ref\affleckeggert{
I Affleck and S Eggert,  Phys. Rev. Lett. 75 (1995) 934.},
local impurity potentials in interacting 1D  electron systems (Quantum
Wires)\ref\kanefisher{C.L. Kane and M.P.A. Fisher,
Phys. Rev. B 46 (1992) 15233} \ref\rwires{E. Wong and I. Affleck, 
Nucl. Phys. B417 (1994) 403.},  
and tunneling point contacts in Fractional Quantum Hall Devices\ref\rFLS{P.
Fendley, A. W. W. Ludwig, and H. Saleur, Phys. Rev. Lett. 74 (1995) 3005;
Phys. Rev. B52 (1995) 8934; Statphys. 19, p.137 (World Scientific, 1996).}.
Notably, for the latter system, the methods of Exact Integrability
as applied to impurities and defects
have recently proven to be a powerful tool for providing
non-perturbative answers to  important strongly interacting
quantum systems, observed  in the Solid State 
laboratory\ref\MillikenWebb{F.P. Milliken,
C.P. Umbach and R.A. Webb, Solid State Commun. 97 (1996) 309.}.
In the 2D statistical mechanics context, the simplest
example is the 2D Ising model in the full plane, with a defect
interaction on the real axis
\ref\rbariev{R. Bariev, Sov. 
Phys. JETP 50 (1979) 613.}\ref\barry{B.
McCoy and J. H. H. Perk, Phys. Rev. Lett. 44 (1980) 840.}\ref\rmussis{G.
Delfino, G. Mussardo and P. Simonetti, Nucl. Phys. B432 (1994) 518.}.  

Generally, a quantum field theory with defect can be formulated
 in terms of 
 an action:
\eqn\eaction{
S_{\rm defect} = \int_{ -\infty < x < \infty}
 dx dt ~ \CL_{\rm bulk} + \lambda \int dt ~
D (0,t) , }
where $D(0,t)$ is a field operator located at the defect at $x=0$.  
The integrability of such theories poses some special problems
in comparison with boundary theories on the half-line $x>0$\ref\rghosh{S. 
Ghoshal and A. Zamolodchikov, Int. J. Mod. Phys. A9 (1994)
3841.}.   If $D$ is a local operator, then the action \eaction\ breaks
translation invariance;  one thus expects the theory to have
non-trivial transmission {\it and} reflection at the defect.   
Unfortunately, the algebraic Yang-Baxter like constraints involving
both transmission and reflection have very limited solutions, and
generally require the bulk theory to be a free field theory\ref\fendleyunp{
P. Fendley, talk given at the conference on `Statistical
Mechanics and Quantum Field Theory', Univ. of Southern California,
16-2- May 1994; and private communication.}\rmussis.
 In particular,  if $D$ defines an integrable
 perturbation of the {\it bulk} CFT, this  does 
not at all imply that 
the corresponding
defect is integrable as well. 
This no-go theorem was circumvented in the works \rwires\rFLS\ 
by exploiting  the following
two special features of the bulk theory\foot{
 consisting  there of free massless
 scalar
fields}.  Namely, if the bulk
theory is a massless conformal field theory (CFT), the defect theory
can be folded onto a boundary field theory  with twice
the central charge (the tensor product of two identical
copies of the original CFT), on the half-line. 
 Secondly, when the bulk theory
consists of  a free scalar field, then the folded boundary theory 
consists of an even and odd combination of the original scalar field
and the odd combination decouples from the boundary. ( Thus, in 
this case,  only  half the central charge of the
tensor product couples to the 
defect after folding, giving
a boundary sine-Gordon model.)
A larger  class of theories that can be treated this     
way  was  studied in \ref\rkonlec{R. Konik and A. LeClair, {\it Purely
Transmitting Defect Field Theories}, hep-th/9703085.}, and can be thought of
as corresponding to  
defect operators that are  purely chiral (left or right-moving);  
in this situation the theory
is purely transmitting in the defect formulation, and the transmission
S-matrices can be mapped onto the reflection S-matrices of the boundary
formulation.  (In some cases, this map requires the
introduction of  defect degrees of freedom.)      

In this paper we consider a general situation
where the bulk theory is a CFT, 
and  the defect operator $D$ is
local  (having both, left and right moving factors).  
By folding the system, 
we show that the class of integrable  defect
theories of this type
 is
 in one-to-one correspondence with integrable bulk perturbations
of {\it two} copies of the CFT.  A large class of such integrable bulk
perturbations was identified in 
\ref\rLLM{A. LeClair, A. Ludwig and G. Mussardo, 
{\it Integrability of Coupled
Minimal Models}, ITP preprint ITP-97-081.}.
 The  resulting  integrable theories 
include defects in minimal models and in coset theories\foot{
These range from
a $c=1$ orbifold, to the level-one current algebra with $c=n$.}
based on $SO(2n)$. 
As opposed to a defect in a theory of free scalar fields (as in
\rwires,\rFLS),
the {\it full} central charge of the tensor product of the original
CFT couples to the defect after folding
in this general situation.
In this paper we  focus in more detail on the case
 when the
bulk  is a $c<1$ minimal unitary CFT and the defect operator is 
the primary field `$\Phi_{1,2}$' or `$\Phi_{2,1}$'.   In the Ising
case this corresponds to taking the spin field  or the
energy operator,  respectively,   as the defect perturbation  
$D$.  For the spin perturbation, this corresponds
to a line of magnetic field in the bulk of the sample.  
 The case of the energy perturbation 
corresponds to a free field theory\foot{ For a massless bulk, this perturbation
is exactly marginal and has been studied by many authors (see e.g.
   \rbariev, \barry, 
\ref\raffleck
{M. Oshikawa and I. Affleck, 
 Phys.Rev.Lett. 77 (1996) 2604; Nucl.Phys. B495 (1997) 533-582.}),
 and the case of a massive bulk was considered
in \rmussis.}.
 In contrast, as described below, 
the spin perturbation 
cannot be solved in the free fermion basis.
 Rather, it   is related  to the
sine-Gordon theory at $\beta^2/8\pi = 1/8$,
 and  a version of the
anisotropic Kondo
model, where these  two cases correspond, as explained
below in more detail, to different choices
of boundary conditions in the ultra-violet (`continuous Neumann'
and `continuous Dirichlet'\raffleck).

 We end this introduction by discussing a general conceptual
 aspect of  integrable massless
renormalization group (RG) flows of defect theories and of  their
corresponding boundary  theories, obtained after folding.
In general, a massless flow between two conformally invariant
boundary conditions on a given CFT in the bulk is characterized
by the following data: (i) the bulk CFT,  (ii) the particular
conformally invariant boundary condition chosen on this CFT before
perturbation  ( i.e. the  ultra-violet limit of the flow), and (iii)
 a particular relevant 
boundary operator chosen to perturb this boundary condition.
Since a given bulk CFT may in general have a large number of conformally
invariant boundary conditions, several flows may be possible\foot{
 In the boundary sine-Gordon model there is only a single boundary
flow, connecting  the only possible conformally
inv. boundary conditions on a free scalar field,  von Neumann and  Dirichlet.} 
(of course,
always consistent with the `g-theorem'\ref\ralud
{I. Affleck and A. Ludwig, Phys. Rev. Lett. 
67 (1991) 161.}).
If two or more of these flows are integrable, then, there must
exist  different reflection matrices, satisfying   the bulk-boundary
Yang-Baxter equations with
 the {\it same} bulk $S$-matrix, corresponding to all the possible
integrable boundary flows. A complete classification of
all these solutions is an open problem. However, the case of a spin-field
defect in an Ising model, analyzed in  section 4,
is precisely an example
of this non-trivial situation (perhaps the first to be understood
completely).
 In this case
 we identify two integrable
flows, connecting boundary conditions with different
ratios $g_{UV}/g_{IR}$ of `ground state degeneracies'
in the ultraviolet and the infrared .
We find two different reflection matrices, one related to
that of the boundary sine-Gordon model, the other related to
the anisotropic Kondo model.

 The paper is organized as follows: in Section 2 we briefly review
basic ideas of integrability in the bulk and at the boundary,
and discuss the  folding procedure as applied to general defect
theories in CFT's. Then we establish that the class of integrable
defect theories is in one-to-one correspondence with integrable
bulk perturbations of the tensor product of two copies a
CFT. In section 3  we apply the general results of
section 2 to the special case of   defect perturbations
of (Virasoro) minimal models. In section 4, we work out in detail
the Ising case (the lowest minimal model). In particular, we 
obtain the bulk $S$-matrices, as well as the boundary reflection
matrices for two different  integrable massless boundary flows, 
and verify that
those satisfy the bulk-boundary bootstrap, and give the correct
values of the boundary entropies.

\newsec{General Aspects of Integrability}

In this section we outline a general strategy for constructing
integrable local defect theories.  We first review some features about
bulk and boundary integrability that we will need.  

\subsec{Bulk and Boundary Integrability}

One can define  integrable bulk theories
as suitable perturbations of conformal field 
theory (CFT)\ref\rzamo{A. Zamolodchikov, Int. J. Mod. Phys. A3 (1988) 743.}: 
\eqn\eIi{
S_{\rm bulk} = S^{\rm CFT}_{\rm bulk}
 + \Lambda \int dx dt ~ \CO(x,t) . }
Here, $S^{\rm CFT}_{\rm bulk}$ denotes a formal action for a specific CFT
in the bulk,
and $\CO$ is a suitably chosen local perturbation making 
\eIi\ integrable.  For many infinite classes of CFT, the integrable
perturbations $\CO$ are known.  For example, for the $c<1$ minimal models,
$\CO$ can be the primary field  $\Phi_{1,3}, \Phi_{1,2}$ or $\Phi_{2,1}$.    

As usual, in Euclidean space let $z=t+ix$, $\zb = t-ix$.  
The integrability of  \eIi\  implies there are an infinite number of
conserved currents $J_L (z), J_R (\zb)$, which are chiral in the 
conformal limit and in the perturbed theory satisfy
\eqn\eIii{
\d_\zb J_L = \d_z H, ~~~~~~~~\d_z J_R = \d_\zb \bar{H} , }
for some $H, \bar{H}$.  
The local operator $\CO (x,t)$ can be factorized into left and right 
moving components in the CFT: 
\eqn\eIiii{
\CO (z, \zb ) =  \CO_L (z)  \CO_R (\zb ) . } 
The conservation laws \eIii\ are a consequence of the fact that for
each $J_L (z)$ the residue of the operator product expansion of
$J_L (z)$ with $ \CO_L (w)$ is a total derivative, and similarly
for ${J}_R$ and $ \CO_R $\rzamo. 

We now turn to boundary theories on the half line $x \geq 0$, with
a boundary interaction at $x=0$.  In the conformal limit, the CFT must
come equipped with a conformally invariant boundary condition 
satisfying $T_L (z) = {T}_R (\zb )$ at $x=0$, where 
$T_L , {T}_R$ are the left and right moving energy momentum tensors. 
This implies that at $x=0$, the left and right moving operators are
identified. 

We make the following claim: {\it If the bulk theory \eIi\ is
integrable, then the boundary theory defined by}:
\eqn\eIiv{
S_{\rm boundary} = S^{\rm CFT}_{\rm bound}
 + \lambda \int dt ~ \L \CO (0,t) }
{\it is also integrable.}\foot{We distinguish between
$\CO_L$ on the infinite plane and its boundary counterpart $\CO^{(L)}$.}
 (Here, $ S^{\rm CFT}_{\rm bound}$ denotes the formal action
of the bulk CFT including the conformally inv. boundary condition.)  
 The reason is that the properties of
the operator product expansion of $J_L (z)$ with $ \CO_L $ described above,
which ensure integrability of the bulk theory, also ensure 
that in the boundary theory one has
\eqn\eIv{
J^{(L)}  (0,t) - \R J (0,t) = - i \d_t \Theta , }
for some $\Theta$, and this implies that a conserved charge $Q$  can be
constructed from $\L J , \R J$:
\eqn\eQ{
Q = \int_0^\infty dx \(\L J + \R J  \) +  \Theta . } 
Using $\d_\zb \L J = \d_z \R J = 0$, one easily sees that 
$\d_t Q =0$.  One can prove \eIv\ using conformal perturbation
theory techniques outlined in\rghosh.  

\subsec{Defect Theories} 

We now describe how to use the above facts to construct integrable
local defect theories.  Consider a defect theory on the full line
$-\infty \leq x \leq \infty$ with a defect at $x=0$. This
can be
formulated as a perturbation of a CFT: 
\eqn\eIvi{
S_{\rm defect} = S^{\rm CFT}_{\rm defect }
 + \lambda \int dt ~ 
D (0,t)  }
 Here $S^{\rm CFT}_{\rm defect }$ denotes
a bulk CFT equipped with a conformally inv. b.c.
at the location of the defect, and
 $D(0,t)$ is an allowed operator at  this defect.
In the  simplest case, discussed below,  $S^{\rm CFT}_{\rm defect }=
S^{\rm CFT}_{\rm bulk}$
is a bulk CFT without any defect at all\foot{
named 
``periodic b.c. '' in \rwires}.
In this case the
perturbing field $D(0,t)$ is an operator of the bulk
theory, placed at the location of the defect.
We now fold the defect theory onto a boundary theory on the 
half line $x\geq 0$.
In general, any bulk
field $\Psi$ can be decomposed into its components 
$\Psi^{(\pm)}$ on either side of the defect:
\eqn\eIvii{
\Psi (x,t) = \p \Psi (x,t) \theta (x) + \m \Psi (x,t) \theta (-x) . }
  Let $\psi_L  (z)$ and $\psi_R  (\zb )$ denote
any left and right-moving fields, respectively,   in the defect CFT, and 
$\psi_L^{(\pm)} (z)$, $\psi_R^{(\pm)} (\zb )$ their components for 
 each side of the defect.  From these we define four boundary
fields in the region $x\geq 0$: 
\eqn\eIviii{\eqalign{
\L \psi_1 (x,t) &= \p{\psi_L} (x,t),~~~~~~~~\R \psi_1 (x,t)=\m{\psi_L} (-x,t)
\cr 
\L \psi_2 (x,t) &= \m{\psi_R} (-x,t) , ~~~~~~~~\R \psi_2 (x,t)=\p{\psi_R} (x,t)
. \cr
}}
The boundary  fields $\L \psi_{1,2}$ are functions of $z = t +ix$,
whereas $\R \psi_{1,2}$ are functions of $\zb$.  
This folding is represented graphically in figure 1.  
\midinsert
\epsfxsize = 5 truein
\bigskip\bigskip\bigskip\bigskip
\vbox{\vskip -.1in\hbox{\centerline{\epsffile{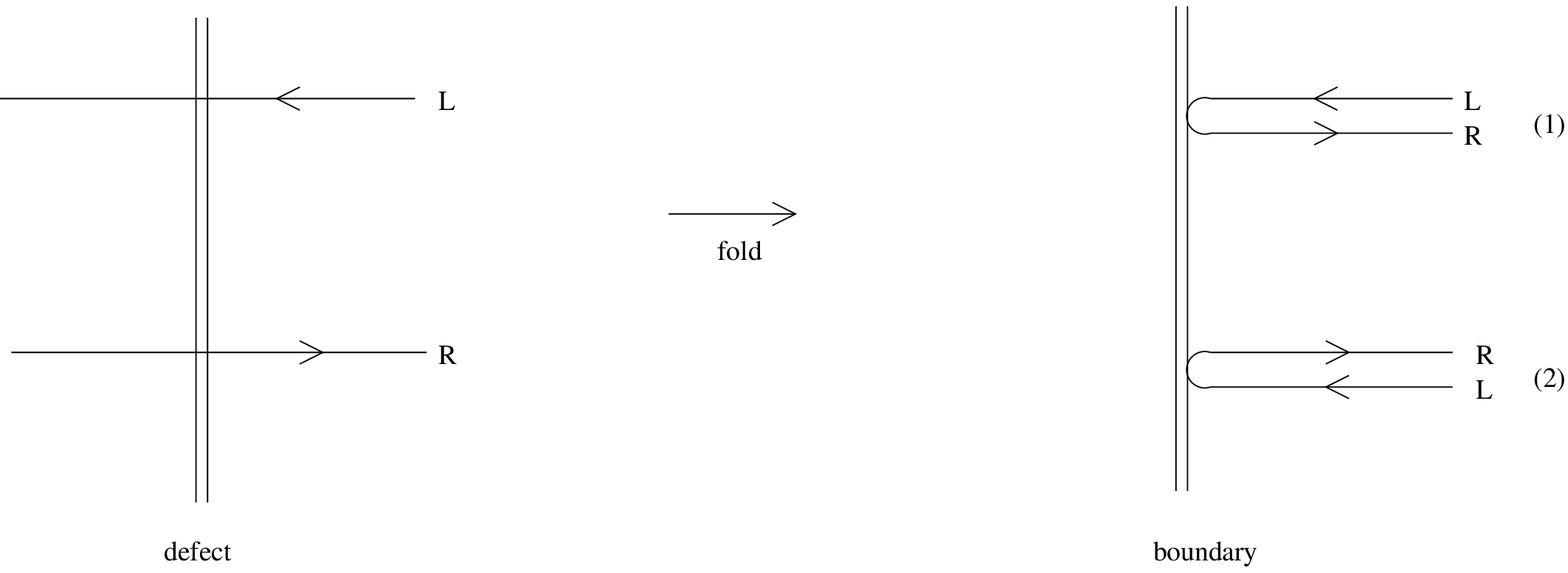}}}
\vskip .1in
{\leftskip .5in \rightskip .5in \noindent \ninerm \baselineskip=10pt
\vskip .1in
~~~~~~~~~~~~~~Figure 1.  Graphical Representation of Folding 
\smallskip}}
\bigskip
\endinsert

For the energy momentum tensor, the defect conformal boundary
condition is 
\eqn\eT{
T^{(-)}_L (0,t) = \p {T_L}  (0,t), ~~~~~~
T_R^{(-)} (0,t) =  T_R^{(+)} (0,t) . }
 These imply the  conformal boundary conditions 
\eqn\eIix{
\L T_1 (z) = \R T_1 (\zb ), ~~~~~~~
\L T_2 (z) = \R T_2 (\zb ), ~~~~~~~ (x=0), }
 individually, on the two copies.
 

In the defect CFT we can factorize the perturbing field 
\eqn\eIx{
D  =  D_L D_R  . } 
To fold the theory, we let $D = (\p D +  \m D )/2$, with 
$D^{(\pm)} = D^{(\pm)}_L D^{(\pm)}_R$, and use the map \eIviii. 
One obtains:
\eqn\eIxi{
S_{\rm boundary} = S^{{\rm CFT}_1
\otimes{\rm CFT}_2
}_{\rm bound} 
+ \frac{\lambda}{2}  \int dt ~ \( \L D_1 \R D_2 + \R D_1 \L D_2 \)  , } 
where $S^{{\rm CFT}_1}_{\rm bound}$, $S^{{\rm CFT}_2}_{\rm bound}$
  denote the  two copies of the original
CFT, with boundary conditions \eIix. 
We further assume that the boundary condition \eIix,\eIviii,\eIvii\
 allows us
to identify $\L D_1 = \R D_1, ~ \L D_2 = \R D_2$ on the boundary;
this is expected to be true up to some possible co-cycles.  One then
obtains
\eqn\eIxi{
S_{\rm boundary} = S^{{\rm CFT}_1\otimes {\rm CFT}_2
}_{\rm bound} + 
 \lambda  \int dt ~  \L D_1 \L D_2  , }
The boundary thus couples the two copies of the CFT. 

 More generally, one may consider the unperturbed theory ($\lambda=0$)
in \eIxi\ to be equipped with {\it  any}  conformally inv. b.c.
on the tensor product of the two copies of the CFT. Tracing back
the steps, this defines  the more general case\foot{ The
perturbation of the `continuous Neumann' boundary condition
in the Ising case discussed below, is an exampe of this situation.} of
a conformally inv. defect theory 
$S^{\rm CFT}_{\rm defect}$
in \eIvi.

Based on the discussion in  section 2.1, we can make the following 
statement.  Let $\CC$ denote the  (bulk)
CFT of the defect theory in 
\eIvi.  {\it Then the defect theory } \eIvi\ {\it is integrable
if the following bulk  perturbation of two copies of $\CC$ 
is integrable}: 
\eqn\eIxii{
S_{\rm bulk} = S^{\CC \otimes \CC}_{\rm bulk} 
+ \Lambda \int dx dt 
~ D_1 D_2 , }
{\it where}
$D_{1,2}$    {\it are the local  fields $D$  from copies
} $1,2$ {\it of}  $\CC$.  

In the situation  when $D$ is purely chiral: $D =  D_L$, or
$ D_R = 1$,  
copy 2 of $\CC$ decouples from the boundary \eIxi.  Thus, in this
situation  if
$ D_L  D_R $ defines an integrable perturbation of {\it one} copy of $\CC$,
the defect theory is integrable.  This is the situation studied in 
\ref\rkon{R. Konik and A. LeClair, {\it Purely Transmitting Defect
Field Theories}, hep-th/9703085.}, and implicit in the works \rwires\rFLS.  

In the more interesting situation where $D$ is local as in 
\eIx, the above requirements are  extremely
 restrictive since they
require known integrable perturbations  of two copies of a CFT. 
Note in particular that if $D$ defines an integrable {\it bulk}
perturbation of one copy of $\CC$, 
then this does not at all ensure that the defect
version is also integrable.   
 Nevertheless, there are large classes
of such integrable perturbations. 
 A first example was provided
 by Vaysburd who showed that two coupled minimal models are 
integrable\ref\rvays{I.
Vaysburd, Nucl. Phys. B446 (1995) 387.}.  
A more general,  and systematic
 scheme for coupling two   (or more)
copies of a conformal
field theory in a way that leads to an integrable theory was described in 
\rLLM, 
and is based on cutting and pasting of Dynkin diagrams 
for the associated affine Toda theories.  This procedure
leads to a large number of new and highly
non-trivial massless  integrable flows in defect theories.

We finish
this section by describing two examples which are limiting 
cases of the integrable defect perturbations of all the minimal models 
considered in the next sections. 
\bigskip
\noindent {\it Ising model in a defect magnetic field.}   
This model is defined by the action 
\eqn\eIxiii{
S_{\rm defect}  = S^{\rm Ising}_{\rm defect}
 + \lambda \int dt ~ \sigma (0,t) , } 
where 
$\sigma$ is the local spin field of dimension $1/8$.  
Two copies of Ising is a $c=1$ orbifold at the radius $R=1$. 
An integrable perturbation of a scalar field is the sine-Gordon
theory.  Thus we expect the boundary version of this theory to be
related to the boundary sine-Gordon theory  at $\beta^2 /8\pi = 1/8$ since
it is  at  this coupling that
the boundary perturbation has dimension
$1/8$.  We will consider this theory in detail below.

\bigskip

\noindent{\it $SU(2)$ Current Algebra at level $1$. } 
This model is defined by 
\eqn\eIxiv{
S_{\rm defect}
 = S_{k=1} + \lambda  \int dt \sum_{m= \pm 1/2} \L {\bar \psi}_m \R \psi_m , }
where $S_{k=1}$ is the $SU(2)$ WZW model at level $1$ and 
$\L \psi_m $ is the primary field in the spinor representation 
with scaling dimension $1/4$.   Here $c=1$.  This current algebra
can be bosonized, thus the integrability follows from the 
usual folding of free bosonic fields.   The folded  $c=2$ theory
is the $SO(4)$ level-one current algebra, and can thus 
 be  formulated as 4 real free fermions.  Since the dimension of
the perturbation is $1/2$, the boundary version of this model is related
to the boundary sine-Gordon theory at the free fermion point.

\newsec{Defect Perturbations of Minimal Models} 

\subsec{The Models}

\def\sigt{\tilde{\sigma}}

We now apply the ideas of the last section to
defect perturbations of the $c<1$ minimal series of unitary CFT. 
We let $\CC_k$ denote the $k$-th minimal model with 
\eqn\ec{
c_k = 1 - \frac{6}{(k+2)(k+3)} , } 
$k=1,2,...$   In $\CC_k$ there exists the local primary fields
$\sigma \equiv \Phi_{1,2}$ and $\sigt \equiv 
\Phi_{2,1}$, with the scaling dimensions 
\eqn\eIIii{\eqalign{
{\rm dim} \( \sigma  \) &= 2 \Delta_\sigma  = 2 \cdot \inv{4} 
\( 1 - \frac{3}{k+3} \)  \cr
{\rm dim} \( \sigt  \) &= 2 \Delta_{\sigt} = 2 \cdot \inv{4} 
\( 1 + \frac{3}{k+2} \) .  \cr
}}
(Here, $\rm dim$ refers to the sum of the left and right conformal
dimensions.)  
We define two defect theories, denoted $\CD_k^{\sigma}$ and 
$\CD_k^{\sigt}$ which are defect  perturbations
of the minimal models by the above operators: 
\eqn\eIIiii{
 S_{\rm defect}   = S^{\CC_k}_{\rm defect}
 + \lambda  \int dt ~ \sigma  (0,t) , }
and similarly with $\sigma \to \sigt$.

\subsec{Integrability} 

Upon folding,  the defect theories $\CD^{\sigma}_k$ become 
boundary theories, which we will denote $\CB^{\sigma}_k$, 
with the action
\eqn\eboundary{
S_{\rm boundary}   = S^{\CC_k \otimes \CC_k}_{\rm bound} 
+ \lambda \int dt 
~ \sigma_1^{(L)} \sigma_2^{(L)} .
}
All of the statements of this section apply with $\sigma \to \sigt$,
and are implied.  
The arguments of the previous section indicate that $ \CD_k^{\sigma}$
are integrable if the  bulk theories which
are defined as bulk perturbations of $\CC_k \otimes \CC_k$ by
$\CO = \sigma_1 \sigma_2$ are integrable.  
We will refer to these bulk theories as $\CM^{\sigma}_k$.

Remarkably, the bulk theories $\CM^{\sigma}_k$  
are in fact integrable\rvays\rLLM.  One way to explain this
is as follows.  The $\CC_k$ minimal model can be
formulated as an $SU(2)$ coset: 
\eqn\eIIvi{
\CC_k = \frac{ SU(2)_k \otimes SU(2)_1}{SU(2)_{k+1} } , }
where 
$SU(2)_k$ is the WZW model at level $k$.  Now we use the fact that 
\eqn\etens{
SU(2)_k \otimes SU(2)_k = SO(4)_k . }
This implies
\eqn\eIIvii{
\CC_k \otimes \CC_k = \frac{ SO(4)_k \otimes SO(4)_1}{SO(4)_{k+1}}. }
As explained by Vaysburd, 
there is an unconventional   way in which to
affinize $SO(4)$, extending the Dynkin diagram by the highest weight
of the vector representation rather than adjoint, leading to the
twisted affine algebra $d^{(2)}_{3}$.\foot{One has the identification
$a_3^{(2)} = d_3^{(2)}$.}   
The spectrum and S-matrices of the bulk theories $\CM_k^{\sigma}$ 
can be obtained as RSOS restrictions of the dual $c_2^{(1)}$ affine
Toda theory which has quantum affine symmetry ${}_q d_3^{(2)}$, as
was done in \rLLM. 

The limiting cases are the Ising model in a defect magnetic field \eIxiii,
which occurs at $k=1$, and the current algebra with defect \eIxiv,
occurring at $k=\infty$.   
It was shown in \rvays\rLLM\ that the bulk theory $\CM_{k=\infty}^{
\sigma}$ 
is equivalent to 4 real massive free fermions, and this is consistent
with our previous remarks for the model \eIxiv.

To solve the folded defect theories $\CB^{\sigma}_k$, 
one must start with the bulk spectrum of particles
that diagonalizes the boundary
interaction;  this spectrum is dictated by the bulk theory 
$\CM^{\sigma}_k$.   Given this spectrum and the bulk 
massless S-matrices, one then finds  boundary reflection
S-matrices that  are consistent with the algebraic constraints
described in the next section.   Alternatively one can think of 
solving the bulk massive theory $\CM_k^\sigma$ with the boundary
interaction of $\CB_k^\sigma$, and then taking the bulk massless limit
$\Lambda \to 0$,  as  
was done for sine-Gordon in
\ref\rfsw{P. Fendley, H. Saleur and N. Warner,
Nucl. Phys. B430 (1994) 577.};  this is more complicated
however since the boundary Yang-Baxter equation is more 
complicated in the massive  versus  massless case.     The bulk S-matrices
for the models $\CM^{\sigma}_k$ are mostly known\rvays\rLLM, 
and in general have an RSOS form.  
In the next section, 
we work out the Ising case where the bulk S-matrix is diagonal.

\newsec{Ising Case}

In this section we work out the Ising case at $k=1$.    The defect problem
is described  by the action \eIxiii.  By the arguments of the last section,
we must first consider  the folded boundary theory
with  the action
\eqn\eIIIi{
S_{\rm bound} = S^{ {\rm Ising}_1\otimes {\rm Ising}_2}_{\rm bound}
+ \Lambda \int_{ x\geq 0} 
{ d x d t} ~ 
\sigma_1 \sigma_2 + \frac{\lambda}{2} 
\int dt ~ \sigma_1^{(L)} \sigma_2^{(L)} , }
where the subscripts refer to copies $1,2$ of the Ising CFT. 
As explained above, the original defect theory corresponds to 
massless bulk term $\Lambda = 0$, and the presence of $\Lambda$ only
serves to determine the bulk spectrum which diagonalizes the boundary
interaction. 

It is important to realize that a theory is not completely defined
by the action \eIIIi;  the theory is only completely specified
once the boundary CFT in the ultra-violet is equipped with 
a conformal boundary condition.  The ${\rm Ising} \otimes 
{\rm Ising}$ CFT is equivalent to a $c=1$ orbifold at radius $1$. 
The possible conformal boundary conditions for the orbifold 
are richer than for the  non-orbifolded theory and were classified in
\raffleck. 
There it was shown that
the  possible boundary conditions are `continuous Neumann' and  
`continuous Dirichlet' depending on the continuous
parameters $\tilde{\phi}_0$ and $\phi_0$ respectively, 
 as well as 
  tensor products of the  known free and fixed  Ising boundary
conditions.
For a
 generic  non-orbifolded scalar field $\phi$,  on the
other hand, the
 only possible conformal
boundary conditions are Neumann and Dirichlet and the dependence
on  zero modes  $\phi_0, \tilde{\phi}_0$ can be removed by a shift 
in $\phi$;  for the orbifold case this is not possible because
of the identification $\phi \sim -\phi$.   

\def\vep{\varepsilon}

The Ising case has additional complexity due to the existence of
exactly marginal bulk and boundary directions which do not  exist for higher
$k$.  Namely, consider adding to the action \eIIIi\ the terms
\eqn\emarg{
\delta S = \Lambda' \int_{ x \geq 0} {d x dt}  ~ 
\varepsilon_1 \vep_2 +
\lambda' \int dt ~ \vep_1^{(L)} \vep_2^{(L)}, }
where $\vep $ is the energy operator with scaling dimension 1. (In terms
of { Majorana  fermions, $\vep = \chi_L \chi_R$.})    
Both $\Lambda'$ and $\lambda'$ have scaling dimension zero,
 and are actually completely marginal.  
The parameter $\Lambda'$ corresponds to moving along the Ashkin-Teller
line of bulk fixed points, and corresponds to a modification of the 
sine-Gordon coupling $\beta$ below.   The parameter $\lambda'$ 
can be shown\raffleck\ to correspond to the 
parameters $\phi_0$ and $\tilde{\phi}_0$
of the `continuous Dirichlet' and  the `continuous Neumann'
 boundary conditions.  
The dimension of the defect operator 
(coupling $\lambda$, in \eIIIi)
 varies continuously with these latter
parameters.  Henceforth we assume that $\Lambda'$ and $\lambda'$ are zero
and consequently the dimension of the defect operator $\sigma$ is 
$1/8$.  Equivalently, the parameters $\phi_0$ and $\tilde{\phi}_0$ 
are taken to be fixed to the values corresponding to $\lambda' = 0$.  

A particular scattering theory describes a flow between two conformal
boundary conditions.  One expects that in the infrared when  
$\lambda \to \infty$ the two copies of Ising each have a fixed
boundary condition, i.e. in the infra-red the boundary condition
is fixed-fixed.   Below we will propose two scattering theories
 which possess
  either the `continuous Neumann' or   the `continuous Dirichlet'
 boundary
condition in the ultra-violet.   We first describe the bulk theory
and the constraints on boundary massless scattering.

\subsec{Bulk Theory}

The bulk theory is a special case of the coupled minimal models 
studied in \rvays\rLLM.  The bulk massive S-matrices are the same as 
for the sine-Gordon theory at $\beta^2/ 8\pi = 1/8$,\foot{The coupling
$\beta$ is normalized in the conventional way where the free fermion
point occurs at $\beta^2 = 4\pi$.} up to some minus signs in the soliton 
sector\rLLM.   These signs can be traced to the fact that 
${\rm Ising} \otimes {\rm Ising}$ is an orbifold CFT of a scalar 
field\ref\rorb{P. Ginsparg, Nucl. Phys. B 295 (1988) 153.}.  
The sine-Gordon theory at this coupling is described by the bulk
action
\eqn\sine{
S = \inv{4\pi} \int dx  dt \(  \inv{2} \( \d_\mu \phi \)^2 
+ \Lambda  \cos \( \phi /2 \) \)  . }
The spectrum consists of two solitonic particles $s_1 , s_2$, 
and 6 breathers with mass ratios 
\eqn\eIIIii{
m_a = 2 m_s  \sin \frac{a\pi}{14} 
,
~~~~~~~~a = 1, 2,..,6. }
In contrast to the sine-Gordon case where the solitons $s_1, s_2$ 
carry $U(1)$ charge $\pm 1$, here  the solitons are not charge conjugates
of each other, but rather are their own anti-particle. 
The bulk S-matrices are 
\eqn\eIIIiii{
S_{s_1 s_1} = S_{s_2 s_2 } = S_{s_1 s_2} = F_{\inv{7}} (\theta) 
F_{\frac{2}{7}} (\theta) F_{\frac{3}{7}} (\theta ) , }
where 
\eqn\eIIIiv{
F_\alpha (\theta ) = \frac{ \tanh \inv{2} ( \theta + i \pi \alpha ) }
{\tanh \inv{2} ( \theta - i \pi \alpha )}
, }
and $\theta $ is the rapidity $E= m\cosh\theta , P = m \sinh \theta$.  
For the breathers one has
\eqn\eIIIv{\eqalign{
S_{ab} (\theta) &= \( \frac{|a-b|}{14} \)
\[ \prod_{k=1}^{ {\rm min} (a,b) -1 }
\( \frac{ |a-b| + 2k}{14} \) \]^2  \( \frac{a+b}{14} \)
\cr
S_{as_1} (\theta) &= S_{a s_2} (\theta) =
(-1)^a \prod_{k=0}^{a-1} \( \frac{7-a + 2k}{14} \),
\cr}}
where
$(\alpha) \equiv F_\alpha (\theta) $. 

The structure of bound states also differs from the sine-Gordon theory;  in
\rLLM\ this structure was obtained from the restricted
${}_q d_3^{(2)}$ symmetry.  
For our problem  only the even breathers ($2,4,6$) are $s_1 - s_1$ 
or $s_2 - s_2$ bound states.  This is to be compared with the sine-Gordon
case where all breathers are $s_1 - s_2$ bound states.

\subsec{Algebraic constraints on massless boundary scattering}

We will need the the algebraic constraints
on the massless boundary scattering matrices.  These can be obtained
by an appropriate massless limit of the equations in \rghosh, as
we now describe.    

In the massive case the boundary
reflection S-matrices $R_a^b$ satisfy the crossing and unitarity 
constraints:
\eqn\eIIIvi{
\eqalign{
& R_a^c (\theta) R_c^b (-\theta ) = \delta_a^b \cr
&K^{ab} (\theta) = S_{a'b'}^{ab} (2\theta) K^{b'a'} (-\theta) , \cr
}}
where $K^{ab} (\theta) = R_{\bar{a}}^b (i\pi /2  - \theta)$.  
The massless limit can be taken by replacing $\theta \to \theta + \alpha$,
and letting $\alpha \to \infty$ and $m\to 0$ keeping $me^\alpha /2 = \mu$ 
held fixed.  This gives the dispersion relation for right movers
$E_a = \mu_a e^\theta$, $P_a = \mu_a e^\theta$, where the $\mu_a$ 
have the same ratios as the bulk masses\eIIIii.   To obtain left-movers,
one lets $\theta \to \theta - \alpha$, and takes the same limits leading
to $E_a = \mu_a e^{-\theta}$, $P_a = - \mu_a e^{-\theta}$.  Thus, given
some reflection S-matrices $R_a^b$ in a massive theory,  we
define the massless scattering matrices as follows:
\eqn\eIIIvii{\eqalign{
\tilde{\CR}_a^b (\theta ) &= \lim_{\alpha \to \infty, m \to 0} 
~ R_a^b (\theta + \alpha) , ~~~~~~{\rm for ~ right ~ movers} \cr
\CR_a^b (\theta ) &= \lim_{\alpha \to \infty, m \to 0} 
~ R_a^b (\theta - \alpha) , ~~~~~~{\rm for ~ left ~ movers. } \cr
}}
The right hand sides of \eIIIvii\ depend on $\theta -\theta_B$ for
right-movers, and $\theta + \theta_B$ for left-movers, 
where $\mu e^{\theta_B}$ is defined as a physical boundary energy 
scale, as described in \rfsw.    
The bulk S-matrix in the massless limit becomes an S-matrix $S_{LL}$
 ($S_{RR}$) for left (right) movers, both of which are the same as 
\eIIIv.  We will also need the `braiding' matrix:
\eqn\eIIIviii{
B_{ab}^{cd} = \lim_{\theta \to - \infty} S_{ab}^{cd} (\theta) . } 
In general $B$ is a solution of the braiding relations.  

The two equations \eIIIvi\ become
\eqn\eIIIix{\eqalign{
\CR_a^c (\theta) \tilde{\CR}_c^b (-\theta) &= \delta_a^b \cr
\tilde{\CR}_{\bar{a}}^b \(i\pi/{2} - \theta \) &= 
B_{cd}^{ab} ~ \CR_{\bar{d}}^c \( i\pi/{2} + \theta \) . \cr
}}
These can be combined into a single equation for $\CR$: 
\eqn\eIIIx{
B_{cd}^{eb} ~ \CR_{a}^{\bar{e}} \( \theta - i\pi/{2} \) 
\CR_{\bar{d}}^c \( \theta + i\pi/{2} \) = \delta_{a}^b . } 
For diagonal bulk scattering, 
\eqn\eIIIxb{
B_{cd}^{ab} = B_{ab} \delta_c^a \delta_d^b , }
one has
\eqn\eIIIxi{
\sum_c  B_{cb} ~ \CR_{a}^{\bar{c}} \( \theta - i\pi/{2} \) 
\CR_{\bar{b}}^c \( \theta + i\pi/{2} \)  = \delta_a^b . } 

Another constraint comes from the boundary bootstrap.  If the particle
of type $c$ is a bound state of particles $a,b$ then the bulk
S-matrix has a pole at $i u_{ab}^c$:
\eqn\epole{
S_{ab}^{a'b'} (\theta) \approx i \frac{f_{ab}^c f_c^{a'b'}}{\theta - i u_{ab}^c}
.}
The fusing angles satisfy $u_{ab}^c + u_{bc}^a + u_{ac}^b = 2\pi$.  
Taking the massless limit of the boundary bootstrap equation in \rghosh\ one
obtains
\eqn\eboot{
f_c^{ab} \CR_d^c (\theta) = 
f_d^{b_1 a_1} \CR_{a_1}^{a_2} (\theta + i \bar{u}_{ac}^b ) 
\CR_{b_2}^b (\theta - i \bar{u}_{bc}^a ) B_{b_1 a_2}^{b_2 a} , }
where $\bar{u} = \pi - u$.  For diagonal bulk scattering, 
\eqn\ebootb{
f_c^{ab} \CR_d^c (\theta) = 
f_d^{b' a'} \CR_{a'}^{a} (\theta + i \bar{u}_{ac}^b ) 
\CR_{b'}^b (\theta - i \bar{u}_{bc}^a ) B_{b' a} . } 

A subset of the fusing angles we used to check our solutions below are the 
following:  
\eqn\efuse{\eqalign{
u_{s_1 s_1}^2 &= \frac{5\pi}{7}, ~~~~~
u_{s_1 s_1 }^4 = \frac{3\pi}{7}, ~~~~~
u_{s_1 s_1 }^6 = \frac{\pi}{7}
\cr
u_{11}^2 &= \frac{\pi}{7}, ~~~~~
u_{22}^4 = \frac{2\pi}{7}, ~~~~~
u_{12}^3 = \frac{3\pi}{14} .\cr }}

\subsec{Boundary reflection S-matrices} 

In our problem the bulk theory has the properties:   
\eqn\eIIIxii{
B_{s_1 s_1} = B_{s_2 s_2} = B_{s_1 s_2 } = 1, ~~~~~~~~~
\bar{s_1} = s_1 , ~ \bar{s_2} = s_2 , } 
whereas for sine-Gordon  $B_{s_i s_j} = -1$ and 
$\bar{s_1} = s_2$.   We now describe two boundary scattering theories
that are both consistent with the above constraints.  

\noindent
{\bf Boundary Sine-Gordon-like Solution ~~~}
In \rfsw\ reflection S-matrices for the boundary sine-Gordon (BSG) theory
were obtained as the massless limit of the results in \rghosh\ref\rghoshb{S.
Ghoshal, Int. J. Mod. Phys. A9 (1994) 4801.}.   
We find 
that by modifying some phases in these reflection S-matrices
we can continue to satisfy \eIIIxi\ with
the new conditions \eIIIxii, and also the bootstrap equation\ebootb.
The result is
\eqn\eIIIxiii{\eqalign{
\CR_{s_1}^{s_2} &= \CR_{s_2}^{s_1} = \frac{e^{-7\theta/2}}{2\cosh 
(7\theta/2 - i\pi/4)} ~ e^{i\delta'} Y(\theta) \cr
\CR_{s_1}^{s_1} &= \CR_{s_2}^{s_2} = \frac{e^{7\theta/2}}{2\cosh 
(7\theta/2 - i\pi/4)} ~ e^{i\delta} Y(\theta) \cr
\CR_{2k} (\theta) &= (-1)^{k-1} \prod_{l=1}^k  F_{\frac{2l-8}{14} }(\theta)\cr
\CR_{2k-1} (\theta) &= i (-1)^{k-1} f_{-\inv{2}} (\theta)
\prod_{l=1}^{k-1} F_{\frac{2l-7}{14}} (\theta) , \cr}}
where for the breathers $\CR_a \equiv \CR_a^a$, and 
\eqn\eIIIxiv{\eqalign{
Y(\theta) &=  F_{- \frac{3}{14} } (\theta ) 
f_{-\inv{2}} (\theta ) , ~~~~~~ 
e^{i\delta} = e^{-i\delta'} = e^{i\pi/4} \cr 
&\cr  f_\alpha (\theta ) &\equiv 
\frac{ \sinh \inv{2} ( \theta + i \pi \alpha ) }
{\sinh \inv{2} ( \theta - i \pi \alpha )}
.\cr
}}
(For the boundary
sine-Gordon theory one has instead $e^{i\delta'} = i, ~ e^{i\delta} =1$.)  
The constraints of crossing-unitarity\eIIIxi\  and the bootstrap\ebootb,
 are easily
checked using $F_\alpha = - f_\alpha f_{1-\alpha} = F_{1-\alpha}$, 
$f_\alpha = f_{\alpha +2}$, $f_\alpha f_{-\alpha} = 1$,
and $f_\alpha (\theta + i\pi \beta) f_\alpha (\theta - i\pi \beta) 
= f_{\alpha - \beta} f_{\alpha + \beta}$.  
 
\noindent
{\bf Kondo-like solution ~~~}
Another solution starts from the minimal one in the soliton sector: 
\eqn\ekond{
\CR_{s_1}^{s_1} = \CR_{s_2}^{s_2} = i f_{-\inv{2}} (\theta) = 
\tanh \inv{2} (\theta - i\pi/2) , ~~~~~ 
\CR_{s_1}^{s_2} = \CR_{s_2}^{s_1} = 0.}
Closing the boundary bootstrap on the breathers using \ebootb\  gives 
\eqn\ekondb{
\CR_a (\theta)  = F_{- a/14 } (\theta ). } 
This is essentially the same as for the anisotropic Kondo model\ref\rfen{P.
Fendley, Phys. Rev. Lett. 71 (1993) 2485.}\ref\rskor{F. Lesage,
H. Saleur and S. Skorik, Nucl. Phys. B474 (1996) 602.},
except that in the latter $R_{s_1}^{s_1} = R_{s_2}^{s_2} = 0.$\foot{We 
 thank
H. Saleur for suggesting this possibility.}.

\subsec{Boundary Entropy} 

In order to determine the ultraviolet (UV) and infra-red (IR) 
fixed points of the  ( massless) flows described by the above
scattering theories, we examine the so-called 
`ground state degeneracies' 
 $g$\ralud.  

Consider the partition function $Z_{\alpha \alpha'}$ on a cylinder
of circumference $L$ and length $R$, with boundary conditions 
$\alpha$  and $\alpha'$ at the ends of the cylinder.  If one formulates
this in a picture where the hamiltonian evolves the system in the direction
along the length of the cylinder, then 
\eqn\eenti{
Z_{\alpha \alpha'} = \langle B_\alpha | e^{-HR} | B_{\alpha'} \rangle , }
where $|B_{\alpha, \alpha'} \rangle$ are boundary states.  In the limit
of large   ratio  $R/ L$, 
\eqn\eentii{
Z_{\alpha \alpha'} = g_\alpha g_{\alpha'} \to
 \langle B_\alpha | 0 \rangle
\langle 0 | B_{\alpha'} \rangle , } 
where $g_\alpha$ is the  `ground state degeneracy'
 for the boundary condition 
$\alpha$.  

For the ${\rm Ising} \otimes {\rm Ising}$ theory, it is known\raffleck that
the `continuous Neumann', `continuous Dirichlet', and `fixed-fixed' conformal
boundary
conditions have  `ground state degeneracies'
$g=\sqrt{2}$, $g=1$, and $g=1/2$ 
respectively.

For the scattering theory the ratio of ultra-violet to 
infra-red  boundary entropies can be computed using the boundary version of
the  thermodynamic 
Bethe ansatz (TBA) \rfsw\ref\rlmss{A. LeClair, G. Mussardo, H. Saleur and
 S. Skorik, Nucl. Phys. B453 (1995) 581.}.  
If the scattering theory is diagonal, the result is 
\eqn\eentiii{
\log \frac{g_{UV}}{g_{IR}} = \sum_a  \inv{2\pi i} \int d\theta  \d_\theta 
\( \log \CR_a \) \log \( 1 + e^{-\varepsilon_a (\theta)} \) , }
where
$\varepsilon_a$ are the bulk TBA pseudo-energies for particle of type $a$,
satisfying the integral equations in \ref\ralzam{Al. Zamolodchikov, Nucl. Phys.
B342 (1991) 695.}.

For our problem, the analysis proceeds much as in \rfsw, where the
flow in the boundary entropy was computed 
for the BSG theory at the reflectionless points. One finds
\eqn\eTBA{
\log \frac{g_{UV}}{g_{IR}} = \sum_{a=1}^6 I^{(a)} \log (1+1/x_a ) 
+ (I^{(+)} + I^{(-)} ) \log (1+ 1/x_\pm ), }
where $x_n$ are related to the constant ultra-violet
values of $\varepsilon_n$, $x_n \equiv \exp( \varepsilon_n (\infty) )$, 
and $I^{(n)} = \int d\theta \d_\theta \CR_n (\theta) /2\pi i$.  
The $I^{(\pm)}$ and $x_\pm$ come from the soliton sector;  
in the BSG case the boundary scattering is diagonal in the basis
$ ( \pm ) = s_1 \pm s_2$.  
The $x_n$ are bulk properties and are known to be 
$x_a = (a+1)^2 -1$ and $x_\pm = 7$.  

It is not difficult
to show that the effect of 
the phase differences in \eIIIxiii\ in comparison to
the BSG case  is merely to distribute the contributions to $g$ from the
solitons differently, but does not modify the sum of the contributions
from $s_1$ and $s_2$.  Namely, $I^{(+)} + I^{(-)} = 7/2$ and
$I^{(a)} = a/2$.  Thus for the BSG-like solution\eIIIxiii\  one has
\eqn\eentiv{
\frac{g_{UV}}{g_{IR}} = \( \frac{8\pi}{\beta^2} \)^{1/2} = 2\sqrt{2} . }
We thus conjecture that this describes the flow between the continuous
Neumann and fixed-fixed boundary condition.  

In the Kondo-like solution\ekond\ekondb, one finds instead 
$I^{(a)} = 1$, $I^{(s_1)} = I^{(s_2)} = 1/2$.  This leads to 
\eqn\eentiv{
\frac{g_{UV}}{g_{IR}} = 2.}
Here, the scattering theory 
is conjectured to describe the flow between a continuous Dirichlet
and fixed-fixed boundary condition. 

Further arguments in favor of these conjectures are as follows.  
Though our bulk theory is an orbifolding of the sine-Gordon theory
which modes out the $U(1)$ symmetry by $Z_2$, this does not
modify significantly the bulk S-matrices, and this suggests
that the boundary reflection S-matrices also have $U(1)$ properties
that are similar to the BSG case.  
It is known that for the non-orbifolded BSG theory the Neumann
boundary condition breaks the $U(1)$ symmetry and this allows
$\CR_{s_1}^{s_2} \neq 0$, as in \eIIIxiii.  On the other hand,
the Dirichlet boundary condition preserves the $U(1)$ symmetry,
as does \ekond.   

 We end with a discussion of the effect of the continuous
parameter characterizing the `continuous Dirichlet' fixed line.
As one moves along this  continuous
line of ultraviolet
fixed points of the flow, parametrized  by $\varphi_0$,
the scaling dimension $\Delta_b(\varphi_0)$ of the perturbing
boundary operator varies continuously with $\varphi_0$.
This is reminiscent of the situation in the anisotropic
Kondo model. Indeed, we propose that the boundary
reflection $S-$matrices are those of the anisotropic
Kondo model at the appropriate value of $\beta$,
namely\foot{
Equation \ekondb is to be replaced with
$ {\cal R}_a(\theta) = F_{-a g/(2-2g)}(\theta)$,
where $g=\beta^2/8 \pi$. Equation \ekond, on the other
hand,
remains unchanged.}
$\beta^2/8 \pi = \Delta_b(\varphi_0)$.
Note that the anisotropic Kondo boundary $S-$matrix leads to a 
ratio of (ground state degeneracy) $g$-values
$g_{UV}/g_{IR} =2$, {\it independent} of the value of $\beta$,
precisely as required for the flow from the `continuous Dirichlet'
fixed line to the  fixed-fixed  Ising boundary condition.

\newsec{Conclusions}

We have described a 
 framework for constructing integrable
massless quantum field theories with local defects which 
generalizes the folding technique  for obtaining
integrable defect theories,  that was previously understood
only for free scalar fields.  The classification
of such theories parallels the classification of integrable bulk
perturbations of two copies of a conformal field theory, as pursued in
\rLLM,  and corresponds to a new class of integrable
theories, obtained by using an analysis of the (extended)
Dynkin diagram of affine Lie algebras. 
 This approach has allowed us to propose a solution to the
problem of an Ising model in a defect magnetic field.   The Ising
case extends to all the minimal models of unitary conformal field theory
with magnetic defects.  The more general class includes also
defects in
  all coset minimal models
based on $SO(2n)$.

\bigskip\bigskip

\centerline{\bf Acknowledgments}

We would like to thank I. Affleck, F. Lesage, G. Mussardo, 
 and H. Saleur for discussions.  
A.Lec.  and A.W.W.L. would  like
to thank the Institute for Theoretical Physics
in Santa Barbara
for the opportunity to carry out this work during the
program 
{\it Quantum Field Theory
in Low Dimensions: From Condensed Matter to Particle Physics}.
 A.W.W.L. also thanks
the Aspen Center for Physics for hospitality.
This research
is supported in part by the National Science Foundation under grant
no. PHY94-07194.  A. Lec. is also supported in part by the National
Young Investigator Program and A. W. W. L. by the A. P. Sloan Foundation.

\listrefs
\end